\documentclass[10pt]{bmc_article}    

\usepackage{cite} 
\usepackage{url}  
\usepackage{ifthen}  
\usepackage{multicol}   
\usepackage[utf8]{inputenc} 
\usepackage{graphicx}
\newcommand{\dgC}{\mbox{$^{\rm{o}}$C}}

\newboolean{publ}

\newenvironment{bmcformat}{\fussy\setboolean{publ}{true}}{\fussy}

\begin{document}
\begin{bmcformat}

\title{Neutral genetic drift can aid functional protein evolution}

\author{Jesse D Bloom$^{1}$%
       \email{Jesse D Bloom - jesse.bloom@gmail.com}%
      \and
         Philip A Romero$^1$%
         \email{Philip A Romero - promero@caltech.edu}
       \and 
         Zhongyi Lu$^1$%
         \email{Zhongyi Lu - lu07@caltech.edu}
       and
         Frances H Arnold\correspondingauthor$^1$%
         \email{Frances H Arnold\correspondingauthor - frances@cheme.caltech.edu}%
      }

\address{%
    \iid(1)Division of Chemistry and Chemical Engineering, California Institute of Technology, Pasadena, CA 91125 USA 
}%

\maketitle

\begin{abstract}

        \paragraph*{Background:} Many of the mutations accumulated by naturally evolving proteins are neutral in the sense that they do not significantly alter a protein's ability to perform its primary biological function.  However, new protein functions evolve when selection begins to favor other, ``promiscuous'' functions that are incidental to a protein's biological role.  If mutations that are neutral with respect to a protein's primary biological function cause substantial changes in promiscuous functions, these mutations could enable future functional evolution. 
      
        \paragraph*{Results:} Here we investigate this possibility experimentally by examining how cytochrome P450 enzymes that have evolved neutrally with respect to activity on a single substrate have changed in their abilities to catalyze reactions on five other substrates.  We find that the enzymes have sometimes changed as much as four-fold in the promiscuous activities.  The changes in promiscuous activities tend to increase with the number of mutations, and can be largely rationalized in terms of the chemical structures of the substrates.  The activities on chemically similar substrates tend to change in a coordinated fashion, potentially providing a route for systematically predicting the change in one function based on the measurement of several others.

        \paragraph*{Conclusions:} Our work suggests that initially neutral genetic drift can lead to substantial changes in protein functions that are not currently under selection, in effect poising the proteins to more readily undergo functional evolution should selection ``ask new questions'' in the future.
\end{abstract}

\ifthenelse{\boolean{publ}}{\begin{multicols}{2}}{}

\section*{Background}
Nature employs proteins for a vast range of tasks, and their capacity to evolve to perform diverse functions is one of the marvels of biology.  Recently, it has become possible to reconstruct convincing scenarios for how new protein functions evolve.  One of the most important conclusions of this work is that the initial steps may occur even before the new functions come under selection~\cite{O'Brien1999,Aharoni2005,Kondrashov2005,Chothia2003,Copley2004,Bridgham2006}.  The reason is that in addition to their primary biological functions, most proteins are at least modestly effective at performing a range of other ``promiscuous'' functions~\cite{O'Brien1999,Aharoni2005,Copley2003,Khersonsky2006,O'Brien2006,O'Loughlin2007}.  In laboratory experiments, selection can rapidly increase these promiscuous functions, often without much immediate cost to a protein's original function~\cite{Aharoni2005}.  In a particularly compelling set of experiments, Tawfik and coworkers have shown that selection for promiscuous activity likely explains the origin and evolution of a bacterial enzyme that hydrolyzes a synthetic compound only recently introduced into the environment~\cite{Aharoni2005,Roodveldt2005,Afriat2006}.  Mounting evidence therefore supports the idea that new protein functions evolve when selection favors mutations that increase an existing weak promiscuous function.\pb
  
But for as long as 50 years, since Linus Pauling and Emile Zuckerkandl published their seminal analysis of molecular change in proteins~\cite{Zuckerkandl1965}, it has been clear that just a small fraction of the mutations that accumulate in naturally evolving proteins are driven by selection for a new function.  Instead, most of the mutations responsible for natural sequence divergence do not change a protein's primary biological function, but rather are due to either neutral genetic drift~\cite{Kimura1983} or pressure for a subtle recalibration of protein properties unrelated to the acquisition of an entirely new function~\cite{Blundell1975}.  However, even though most mutations accumulate under the constraint that they not interfere with a protein's primary function, they could still substantially alter other, promiscuous functions.  Such alterations could then aid in the subsequent evolution of new functions.\pb

Here we have experimentally investigated this possibility using a set of enzymes that have undergone genetic drift that is neutral with respect to a well-defined laboratory selection criterion for enzymatic activity on a single substrate~\cite{Bloom2007c}.  We have examined how these enzymes have changed in their promiscuous activities on five other substrates.  As described below, we find that the enzymes have often undergone substantial changes in their promiscuous activities, suggesting that neutral genetic drift could play an important role in enabling future functional evolution.\pb

\section*{Results and Discussion}
\subsection*{A set of neutrally evolved cytochrome P450 enzymes}
We focused our analysis on cytochrome P450 proteins.  P450s are excellent examples of enzymes that can evolve to catalyze new reactions, since they are involved in a wide range of important functions such as drug metabolism and steroid biosynthesis~\cite{Montellano1995,Lewis2001}.  We worked with P450 BM3, a cytosolic bacterial enzyme that catalyzes the subterminal hydroxylation of medium- and long-chain fatty acids~\cite{Munro2002}.  We have previously described a set of P450 BM3 heme domain variants that were created by laboratory neutral evolution from a common parent sequence~\cite{Bloom2007c}.  Here we briefly recap the procedure used to create these P450s in order to explain their origin and why they can properly be viewed as the product of neutral genetic drift.\pb

The essential difference between neutral genetic drift and adaptive evolution is that in the former case mutations that have no substantial effect on fitness spread stochastically in a population, while in the latter case mutations spread because they are beneficial and so favored by selection.  Of course, it may be difficult to discern whether a specific mutation in a natural population has spread neutrally or due to favorable selection.  But in the laboratory it is possible to define an arbitrary selection criterion to ensure that all mutations spread due to neutral genetic drift.  Specifically, we imposed the requirement that the P450s had to hydroxylate the substrate 12-$p$-nitrophenoxydodecanoic acid (12-pNCA) with an activity exceeding a specific threshold~\cite{Bloom2007c}.  All mutant P450s were therefore straightforwardly classified as either functional (if they exceeded the threshold) or nonfunctional (if they did not).  While this selection criterion is obviously a simplification of natural evolution, we believe that for the current purpose it is a reasonable abstraction of the evolutionary requirement that an enzyme's primary activity exceed some critical level in order to allow its host organism to robustly survive and reproduce.  To implement laboratory neutral evolution using this selection criterion, we began with a single parent P450 BM3 heme domain variant (called R1-11) and used error-prone PCR to create random mutants of this parent~\cite{Bloom2007c}.  Mutants that failed to yield sufficient active protein to hydroxylate  at least 75\% of the 12-pNCA of the R1-11 parent when expressed in \textit{Escherichia coli} were immediately eliminated, while all other mutants were carried over to the next generation with equal probability.  Any mutations that spread among the offspring sequences were therefore by definition due to neutral genetic drift, since there was no opportunity for any functional mutant to be favored over any other.  We emphasize that the fact that the mutations spread due to neutral genetic drift does not mean that they have no effect on the protein's properties.  Indeed, one of the growing realizations about protein evolution is that mutations that spread by neutral genetic drift may still have an impact on future evolution~\cite{DePristo2005,Bloom2007}.  One mechanism for this impact is that neutral genetic drift can change a protein's stability and so alter its tolerance to future mutations~\cite{Bloom2005,Besenmatter2007,Bloom2006}.  As will be demonstrated below, another mechanism is that neutral genetic drift can alter a protein's promiscuous functions.\pb

As described previously~\cite{Bloom2007c}, the end result of the neutral evolution was 44 different P450 variants, each of which satisfied the selection criterion for activity on 12-pNCA (these are the combined final sequences from the monomorphic and polymorphic populations in \cite{Bloom2007c}).  For the current study, we analyzed the promiscuous activities of 34 of these neutrally evolved P450 variants.  The sequence diversity of these P450s is shown in the phylogenetic tree of Figure \ref{fig:tree}; they have accumulated an average of four nonsynonymous mutations each. \pb

\subsection*{Activities of the neutrally evolved P450 enzymes}
All of the P450 variants had evolved under selection solely for their ability to hydroxylate 12-pNCA.  We examined their promiscuous hydroxylation activities on the five other substrates shown at the top of Figure \ref{fig:heatmap}.  Two of these promiscuous substrates, propranolol and 2-amino-5-chlorobenzoxazole (also known as zoxazolamine), are drugs that are metabolized by human P450s~\cite{Otey2005,Lasker1982}.  The other three promiscuous substrates, 11-phenoxyundecanoic acid, 2-phenoxyethanol, and 1,2-methylenedioxybenzene, are organic compounds of increasing structural dissimilarity to 12-pNCA.  The parent P450 possessed at least some hydroxylation activity on all of these substrates (throughout the remainder of this work, ``activity'' refers to total substrate turnovers per enzyme).    \pb

We measured the activities of all 34 neutrally evolved P450s on the five promiscuous substrates as well as 12-pNCA.  Figure \ref{fig:heatmap} shows the fold change in activity of each of the variants relative to the parent P450 on all six substrates, and Figure \ref{fig:foldchanges} shows the same data with standard errors.  As is apparent from these figures, many of the neutrally evolved P450s have undergone changes in their activities that substantially exceeded the standard errors of the measurements.  Even on 12-pNCA, some of the variants have undergone modest increases or very mild decreases in activity .  The modest increases in 12-pNCA activity were unsurprising, since the parent P450 only hydroxylates 12-pNCA with about a quarter of the activity reported for a P450 engineered for maximal 12-pNCA activity~\cite{Cirino2003}.  Likewise, the mild decreases in 12-pNCA activity were due to the fact that during neutral evolution the P450s were only required to maintain this activity above a minimal threshold (75\% of the total 12-pNCA conversion of the parent protein when expressed in \textit{E. coli}~\cite{Bloom2007c}).  The changes in the promiscuous activities were often much larger than those on 12-pNCA.  For example, several of the neutrally evolved variants have undergone nearly four-fold increases in activity on one or more of 2-phenoxyethanol, 2-amino-5-chlorobenzoxazole, and 1,2-methylenedioxybenzene.  Other variants have experienced equally large decreases in one or more of the promiscuous activities.\pb

\subsection*{Broad patterns of change in activity can be rationalized in terms of substrate properties}
The data in Figures \ref{fig:heatmap} and \ref{fig:foldchanges} clearly indicate that some of the P450s have undergone substantial changes in their activities.  In an effort to understand the nature of these changes, we sought to determine whether there were any clear patterns in the activities.  In Figure \ref{fig:heatmap}, the substrates have been hierarchically clustered so that each successive cluster contains substrates on which the P450s have increasingly similar activities (the clustering is illustrated by the tree-like dendrogram at the top of the figure, with similar substrates in adjacent columns).  The clustering of the substrates is readily rationalized in terms of their chemical structures.  For example, 2-amino-5-chlorobenzoxazole and 1,2-methylenedioxybenzene cluster, meaning that P450s with high activity on one of these substrates also tend to have high activity on the other.  Presumably, they cluster because the similarity of their structures (both are fusions of six and five membered rings) means that they have similar modes of docking in the substrate binding pocket.  Likewise, 12-pNCA and 11-phenoxyundecanoic acid are phenoxycarboxylic acids of similar chain length, and are in the same cluster.  To a lesser extent, 2-phenoxyethanol resembles 12-pNCA and 11-phenoxyundecanoic acid in its phenolic ether structure, and it falls into a higher level cluster with these two substrates.  Propranolol shares a fused ring structure with 2-amino-5-chlorobenzoxazole and 1,2-methylenedioxybenzene, and these three substrates share a common higher level cluster.  Overall, the hierarchical clustering indicates that substrates that appear similar to the human eye are also ``seen'' this way by the P450s, since the P450s tend to increase or decrease their activities on these substrates in a coordinated fashion.\pb

Figure \ref{fig:heatmap} also shows the P450 variants arranged in hierarchical clusters.  A visual inspection immediately indicates that there is an overall association among all of the activities.  Some of the P450 variants (redder rows) tend to show improved activity on most substrates, while others (bluer rows) tend to show decreased activity on most substrates.  Taken together with the clustering of the similar substrates, this overall association suggests that there are two main trends in the activity changes.  First, the P450s appear to have undergone general changes in their catalytic abilities that are manifested by broad increases or decreases in activity on all substrates.  Second, the P450s appear to have experienced shifts in specificity to favor either the fused ring or the phenolic ether substrates.\pb

To test whether these two apparent trends in activity changes are supported by a quantitative examination of the data, we performed principal component analysis.  Principal component analysis is a well-established mathematical technique for finding the dominant components of variation in a data set, essentially by diagonalizing the covariance matrix.  As suggested by the foregoing visual inspection, principal component analysis revealed that two components explained most of the changes in P450 activity (Table \ref{tab:pca}).  The first component contained positive contributions from all six substrates, and so represents a general improvement in catalytic ability.  The second component contained positive contributions from the fused ring substrates and negative contributions from the phenolic ether substrates, and so represents an increased preference for the former class of substrates over the latter.  Together, these two components explain 82\% of the variance in activities among the 34 P450 variants.  The remaining 18\% of the variance is explained by the four remaining components, which represent more subtle shifts in activity that are less easily rationalized with intuitive chemical arguments.\pb
  
\subsection*{Overall distributions of change in the activities}
The preceding sections have demonstrated that neutral genetic drift can lead to substantial changes in P450 activities, and that many of these changes can be understood as resulting from either fairly general increases/decreases in catalytic ability or shifts in preference for different broad classes of substrate structures.  In this section, we examine whether there are any pervasive trends in the distributions of activity changes --- for example, did most of the promiscuous activities tend to increase or decrease?  If a property is not under any evolutionary constraint, then during neutral genetic drift its values might be expected to be distributed in a roughly Gaussian fashion, as the neutrally evolving proteins freely sample from the presumably normal underlying distribution.  On the other hand, if a property is constrained by selection to remain above a certain threshold, then during neutral genetic drift its values should display a truncated distribution since selection culls proteins with values that fall below the threshold (such a distribution has been predicted for protein stability by simulations~\cite{Taverna2002} and theory~\cite{Bloom2007}).   \pb

Figure \ref{fig:changedistribution} shows the distribution of changes in activity for each of the six substrates.  The distribution for 12-pNCA appears to be truncated on the left, as expected since the P450s neutrally evolved under a requirement to maintain the ability to hydroxylate 12-pNCA.  Some of the P450s have undergone a mild decrease in 12-pNCA activity, reflective of the fact that the neutral evolution selection criterion provided a small amount of latitude by allowing the total amount of hydroxylated 12-pNCA to drop to 75\% of the parental value~\cite{Bloom2007c}.  A number of P450s have neutrally evolved 12-pNCA activity that modestly exceeds that of the parent --- again unsurprising, because the parental 12-pNCA activity falls well below the maximal value achievable for this type of protein~\cite{Cirino2003}.  The distribution for 11-phenoxyundecanoic acid resembles that for 12-pNCA, probably because activities on these two chemically similar substrates are highly linked, as discussed in the previous section.\pb

The other four promiscuous activities are less linked to 12-pNCA activity, and their distributions are much more symmetric.  The symmetric shapes of these distributions suggest that neutral genetic drift has sampled from a roughly Gaussian distribution for these four promiscuous activities.  For three of the substrates (propranolol, 2-amino-5-chlorobenzoxazole, and 1,2-methylenedioxybenzene), the distributions of activities are approximately centered around the parental activity.  This centering indicates that the promiscuous activities of the parent on these three substrates are typical of what would be expected of a neutrally evolved P450.  The distribution for 2-phenoxyethanol, on the other hand, is shifted towards activities higher than that of the parent.  This shift indicates that the parent is less active on 2-phenoxyethanol than a typical neutrally evolved P450.\pb

If the activity distributions of Figure \ref{fig:changedistribution} truly reflect what would be expected after a very long period of neutral genetic drift (i.e., if they are ``equilibrium'' distributions), then each variant represents a random sample from the underlying distribution of activities among all P450s that can neutrally evolve under this selection criterion.  In this case, there should be no correlation between the extent of change in activity and the number of accumulated mutations, since the P450s should have lost all ``memory'' of the parent's activity.  On the other hand, if there has not been enough neutral genetic drift to completely eliminate residual memory of the parent's activity, then variants with fewer mutations should more closely resemble the parent's activity profile.  To test whether the activity distributions of the P450 variants had equilibrated, we computed the correlation between the magnitude of each variant's change in activity and the number of nonsynonymous mutations it possessed relative to the parent.  Table \ref{tab:mutcorrelation} shows that the magnitude of activity change is positively correlated with the number of mutations for all six substrates.  Although the correlations for the individual substrates are mostly not statistically significant due to the small number of samples, the overall correlation for all six substrates is highly significant ($P = 10^{-3}$).  Therefore, the P450 activities are still in the process of diverging from the parental values by neutral genetic drift.  If the variants were to undergo further neutral genetic drift, we would expect to see even larger changes in their promiscuous activities.\pb

We also examined whether P450 variants with mutations near the substrate binding pocket were more likely to have undergone large changes in their activities.  Five of the P450 variants had a mutation to a residue that was within 5 \AA\  of the surrogate substrate in the P450 BM3 crystal structure~\cite{Haines2001}: variant M2 had A74V, M8 had A330V, M13 had M354I, M15 had A74P, and M24 had I263V~\cite{Bloom2007c}.  Two of these mutated residues are of clear importance, since mutating residue 74 has previously been shown to shift substrate specificity~\cite{Li2001,Li2001b,Otey2005} and residue 263 plays a role in the substrate-induced conformational shift~\cite{Pylypenko2004}.   We compared the activity changes for the five variants with mutations near the binding pocket to those for the 29 variants without any such mutations, computing the magnitude of activity change as the absolute value of the logarithm (base two) of the fold change in activity averaged over all six substrates.  The average magnitude of activity change for the five variants with mutations near the active site was 0.88, while the average for the other 29 variants was 0.47.  These averages are significantly different, with an unequal variance T-test $P$-value of $10^{-2}$.  Therefore, variants with mutations near the substrate binding pocket are especially likely to have altered activities, although many variants without mutations near the pocket also underwent substantial activity changes.  \pb

\section*{Conclusions}
We have shown that neutral genetic drift can lead to changes of as much as four-fold in the promiscuous activities of P450 proteins.  The ubiquity of these changes is striking --- even though many of the neutrally evolved P450s had only a handful of mutations, most of them had experienced at least some change in their promiscuous activities.  P450s may be especially prone to this type of change, since their catalytic mechanism involves large substrate-induced conformational shifts~\cite{Modi1996} that can be modulated by mutations distant from the active site~\cite{Glieder2002,Meinhold2005,Li2001}.  In addition, P450s have a tendency to eventually undergo irreversible inactivation that can be promoted by reduced coupling between substrate binding and conformational shifts, as well as by other poorly understood determinants of catalytic stability~\cite{Munro2002,Loida1993,Bernhardt2006}.  There are therefore ample opportunities for mutations that spread by neutral genetic drift to cause subtle alterations in a P450's promiscuous activities.  But we believe that neutral genetic drift is also likely to cause substantial changes in the promiscuous activities of enzymes with other catalytic mechanisms.  In support of this idea, a recent study by Tawfik and coworkers~\cite{Amitai2007} indicates that mutations with little effect on the native lactonase activity of serum paraoxonase can alter this enzyme's promiscuous activities.  Taken together, this study and our work suggest that neutral genetic drift allows for changes in promiscuous protein functions.  These changes could in turn have important implications for future functional evolution.  For example, one can easily imagine a scenario in which neutral genetic drift enhances a promiscuous protein function, and then a subsequent gene duplication allows natural selection to transform one of the genes into the template for a protein with a full-fledged new functional role~\cite{O'Brien1999,Aharoni2005,Kondrashov2005,Chothia2003,Copley2004,Bridgham2006}.\pb

One of the most attractive aspects of our study is the degree to which the changes in P450 activities during neutral genetic drift could be understood in terms of the chemical structures of the substrates.  Neutral genetic drift did not simply cause unpredictable shifts in activities.  Instead, most of the variation was explained by two eminently intuitive components: an overall increase or decrease in catalytic ability, and a preference for either fused ring or phenolic ether substrates.  We have suggested that neutral genetic drift under a fixed selection criterion can be viewed as sampling underlying ``equilibrium'' distributions of activities.  The distributions for different activities are linked, since we have shown that P450s with good activity on one substrate will frequently also be highly active on chemically similar substrates (similar linkages have been observed in P450s created by recombination~\cite{Landwehr2007}).  So while it may be impossible to know exactly how any specific mutation will affect a given activity, measuring a handful of activities allows one to make relatively accurate predictions about other closely linked activities.  The prerequisite for making such predictions is an understanding of the linkages among activities in the set of sequences explored by neutral genetic drift (the neutral network).  We have made the first steps in elucidating these linkages for P450s that have neutrally evolved under one specific selection regime.  The linkages are very similar to those that would have been made by an organic chemist grouping the substrates on the basis of their chemical structures.  Knowledge of these linkages is of use in understanding the origins of enzyme specificity~\cite{O'Loughlin2007,Varadarajan2005} --- if an enzyme displays high activity on one substrate but low activity on another, then either these two activities are negatively linked during neutral genetic drift or selection has explicitly disfavored one of them.\pb

Our work also has implications for the general relationship between neutral genetic drift and adaptive evolution.  A number of studies focused on RNA~\cite{Huynen1996,Huynen1996b,Fontana1998} or computational systems~\cite{vanNimwegen1997,vanNimwegen2000} have suggested that genetic drift might aid in adaptive evolution.  Our study and that of Tawfik and coworkers~\cite{Amitai2007} support this notion for the evolution of new protein functions.  However, the way that drift in promiscuous functions promotes adaptive evolution is slightly different than the paradigm proposed for RNA~\cite{Huynen1996,Huynen1996b,Fontana1998} and computational systems~\cite{vanNimwegen1997,vanNimwegen2000}.  In those systems, neutral genetic drift is envisioned as allowing a sequence to move along its neutral network until it reaches a position where it can jump to a new higher-fitness and non-overlapping neutral network.  In contrast, promiscuous protein functions change even as a protein drifts along a single neutral network.  The adaptive benefits of this drift come when new selective pressures suddenly favor a previously irrelevant promiscuous function, in effect creating a new neutral network that overlaps with parts of the old one.\pb

Overall, experiments have now demonstrated two clear mechanisms by which neutral genetic drift can aid in the evolution of protein functions.  In the first mechanism, neutral genetic drift fixes a mutation that increases a protein's stability~\cite{Serrano1993,DePristo2005,Bloom2007}, thereby improving the protein's tolerance for subsequent mutations~\cite{Bloom2005,Besenmatter2007,Bloom2006}, some of which may confer new or improved functions~\cite{Bloom2006}.  In the second mechanism, which was the focus of this work and the recent study by Tawfik and coworkers~\cite{Amitai2007}, neutral genetic drift enhances a promiscuous protein function.  This enhancement poises the protein to undergo adaptive evolution should a change in selection pressures make the promiscuous function beneficial at some point in the future.  \pb

\section*{Methods}
\subsection*{Determination of P450 activities}
We attempted to determine the activities of all 44 neutrally evolved P450 variants described in \cite{Bloom2007c} (22 from the final monomorphic populations and 22 from the final polymorphic population).  Ten of these variants expressed relatively poorly in the procedure used here (as described in more detail below), and so were eliminated from further analysis since their low expression led to large errors in the activity measurements.  That left activity data for the 34 neutrally evolved P450 variants listed in Figures \ref{fig:heatmap} and \ref{fig:foldchanges}, as well as for the R1-11 neutral evolution parent.  The activities for each of these P450 variants were measured on all six substrates (12-pNCA, 2-phenoxyethanol, propranolol, 11-phenoxyundecanoic acid, 2-amino-5-chlorobenzoxazole, and 1,2-methylenedioxybenzene).  In all cases, the activities represent the total amount of product produced after two hours, and so are in units of total turnovers per enzyme.  P450 BM3 enzymes typically catalyze only a finite number of reaction cycles before becoming irreversibly inactivated, and we believe that all reactions were essentially complete after two hours, so these activities should represent the total turnovers of the enzymes during their catalytic lifetimes.\pb

To obtain P450 protein for the activity measurements, we expressed the protein using catalase-free \textit{Escherichia coli}~\cite{Barnes1991} containing the encoding gene on the isopropyl $\beta$-D-thiogalactoside (IPTG) inducible pCWori~\cite{Barnes1991} plasmid (the catalase is removed since it breaks down the hydrogen peroxide used by the P450).  The sequences of the P450 variants are detailed in \cite{Bloom2007c}.  We used freshly streaked cells to inoculate 2 ml cultures of Luria Broth (LB) supplemented with 100 $\mu$g/ml of ampicillin, and grew these starter cultures overnight with shaking at 37\dgC.  We then used 0.5 ml from these starter cultures to inoculate 1 L flasks containing 200 ml of terrific broth (TB) supplemented with 100 $\mu$g/ml of ampicillin.  The TB cultures were grown at 30\dgC\ and 210 rpm until they reached an optical density at 600 nm of $\approx$0.9, at which point IPTG and $\delta$-aminolevulinic acid were added to a final concentration of 0.5 mM each.  The cultures were grown for an additional 19 hours, then the cells were harvested by pelletting 50 ml aliquots at 5,500 g and 4\dgC\ for 10 min, and stored at -20\dgC.  To obtain clarified lysate, each pellet was resuspended in 8 ml of 100 mM [4-(2-hydroxyethyl)-1-piperazinepropanesulfonic acid] (EPPS), pH 8.2 and lysed by sonication, while being kept on ice.  The cell debris was pelleted by centrifugation at 8,000 g and 4\dgC\ for 10 minutes, and the clarified lysate was decanted and kept on ice. \pb

To perform the assays, various dilutions of the clarified lysate were used to construct a standard curve.  For each sample, we prepared dilutions of the clarified lysate in the 100 mM EPPS (pH 8.2) buffer to create samples for the standard curves.  The dilutions were 100\% clarified lysate (undiluted), 67\% lysate, 40\% lysate, 25\% lysate, 17\% lysate, 10\% lysate, 6.7\% lysate, and 4.0\% lysate.  Similar dilutions were also prepared of the clarified lysate of \textit{E. coli} cells carrying a null pCWori plasmid in order to assess the background readings from lysate without any P450.  A pipetting robot was then used to dispense 80 $\mu$l of this series of clarified lysate dilutions into 96-well microtiter plates.  Duplicate microtiter plates were then assayed for P450 concentration and total enzymatic activity on each of the six substrates.  The R1-11 parent was assayed four times rather than in duplicate.  To minimize variation, all of these assays were performed in parallel, with the same stock solutions, and on the same day.\pb

The P450 concentration was determined using the carbon monoxide (CO) difference spectrum assay~\cite{Otey2003}.  Immediately before use, we prepared a 5$\times$ stock solution of 50 mM sodium hydrosulfite in 1.3 M potassium phosphate, pH 8.0.  A multichannel pipette was used to add 20 $\mu$l of this stock solution to each well of the microtiter plates (which contained 80 $\mu$l of a dilution of clarified lysate), so that the final sodium hydrosulfite concentration was 10 mM in each well.  The plates were briefly mixed and the absorbances were read at 450 and 490 nm.  The plates were then incubated in a CO binding oven~\cite{Otey2003} for 10 minutes to bind CO to the iron.  The absorbance was then again read at 450 and 490 nm.  The amount of P450 is proportional to the increase in the magnitude of the absorbance at 450 nm minus the absorbance at 490 nm.  At each dilution along the standard curve, the reading for the null control (lysate dilutions without P450) was subtracted from the reading for each P450 variant to control for clarified lysate background.  Additional file \ref{add:readings} shows the standard curves for all P450 variants.  Ten P450 variants had standard curve slopes less than or equal to 0.020, indicating a low P450 concentration.  These were the ten P450 variants that we discarded from further analysis, since the low P450 concentration decreased the accuracy of the measurements.\pb

To determine the activity on 12-pNCA, we monitored the formation of the yellow 4-nitrophenolate compound that is released upon hydroxylation of the twelfth carbon in the 12-pNCA molecule~\cite{Schwaneberg1999,Cirino2003}.  Immediately before use, we prepared a 6$\times$ stock solution of 12-pNCA by adding 3.6 parts of 4.17 mM 12-pNCA in DMSO to 6.4 parts 100 mM EPPS, pH 8.2.  A multichannel pipette was used to add 20 $\mu$l of this stock solution to each well of the microtiter plates (which contained 80 $\mu$l of a dilution of clarified lysate).  The plates were briefly mixed, and the absorbance was read at 398 nm.  To initiate the reactions, we then prepared a 6$\times$ stock solution of 24 mM hydrogen peroxide in 100 mM EPPS, pH 8.2, and immediately added 20 $\mu$l of this solution to each well of the microtiter plate and mixed.  The final assay conditions were therefore 6\% DMSO, 250 $\mu$M 12-pNCA, and 4 mM hydrogen peroxide.  The reactions were incubated on the benchtop for two hours, and the total amount of enzymatic product was quantified by the gain in absorbance at 398 nm.  At each dilution along the standard curve, the corresponding null control lysate dilution  was subtracted from the reading to control for lysate background.  Additional file \ref{add:readings} shows the standard curves for all P450 variants.\pb

The activities on 2-phenoxyethanol, propranolol, 11-phenoxyundecanoic acid, 2-amino-5-chlorobenzoxazole, and 1,2-methylenedioxybenzene were determined using the 4-aminoantipyrene (4-AAP) assay~\cite{Otey2003b,Otey2004}, which detects the formation of phenolic compounds.  For each of these five substrates, immediately before use we prepared a 6$\times$ substrate stock solution.  These stock solutions were 6\% DMSO and 6\% acetone in 100 mM EPPS, pH 8.2, with an amount of substrate added so that the substrate concentrations in the stock solutions were: 150 mM for 2-phenoxyethanol, 30 mM for propranolol, 5 mM for 11-phenoxyundecanoic acid, 12 mM for 2-amino-5-chlorobenzoxazole, and 120 mM for 1,2-methylenedioxybenzene.  The stock solutions were prepared by first dissolving the substrate in the DMSO and acetone, and then adding the EPPS buffer.  In some cases, the stock solution became cloudy upon addition of the buffer, but there was no immediate precipitation, so we could still pipette the stock solution.   A multichannel pipette was used to add 20 $\mu$l of the appropriate substrate stock solution to each well of the microtiter plates (which contained 80 $\mu$l of a dilution of clarified lysate).  To initiate the reactions, we then added 20 $\mu$l of the freshly prepared 6$\times$ hydrogen peroxide stock solution (24 mM hydrogen peroxide in 100 mM EPPS, pH 8.2) and mixed.  We incubated the plates on the benchtop for two hours.  To detect the formation of phenolic products, a pipetting robot was used to add  and mix 120 $\mu$l of quench buffer (4 M urea in 100 mM sodium hydroxide) to each well.  We then used the robot to add and mix 36 $\mu$l per well of 0.6\% (w/v) of 4-aminoantipyrene in distilled water, and immediately read the absorbance at 500 nm.  To catalyze formation of the red compound produced by coupling a phenolic compound to 4-aminoantipyrene~\cite{Otey2003b,Otey2004}, we then used the pipetting robot to add and mix 36 $\mu$l per well of 0.6\% (w/v) of potassium peroxodisulfate in distilled water.  The plates were incubated on the benchtop for 30 minutes, and the amount of product was quantified by the gain in absorbance at 500 nm.  At each dilution along the standard curve, the corresponding null control lysate dilution  was subtracted from the reading to control for lysate background.  Additional file \ref{add:readings} show the standard curves for all P450 variants.\pb
 
In order to extract enzymatic activities from the standard curves, we fit lines to the data points.  For some of the substrates (most notably 12-pNCA and 2-phenoxyethanol), many of the P450 variants were sufficiently active to either saturate the substrate or exceed the linear range of absorbance readings.  Therefore, we examined each standard curve by eye to determine which points remained in the linear range.  Lines were then fit to the points in the linear range.  These fits are shown in Additional file \ref{add:readings}.  In the plots in this file, all points that were deemed to fall in the linear range (and so were used for the fits) are shown as filled shapes, while all points that were deemed to fall outside the linear range (and so were not used in the fits) are shown as empty shapes.  The figures show the slopes of the lines for all replicates (two replicates for all P450 variants except for R1-11, which had four replicates).  These slopes are averaged for a best estimate of the slope, and the standard error computed over these two measurements is also reported.  \pb

To compare the activities (total substrate turnovers per enzyme) among the different P450 variants, it is first necessary to normalize to the enzyme concentration.  To do this, we took the ratio of the slope for each substrate divided by the slope of the CO different spectrum, propagating the errors.  These normalized slopes are proportional to the activity on each substrate.  The normalized slopes are given in Additional file \ref{add:activity_data}.  This file also lists the number of nonsynonymous mutations that each P450 variant possesses relative to the R1-11 parent sequence, as originally reported in \cite{Bloom2007c}.  These normalized slopes allow for accurate comparisons among the P450 variants, and were used in the analyses in this paper.  To convert these normalized slopes into total substrate turnovers per enzyme, it is necessary to multiply them by the ratio of extinction coefficients.  The extinction coefficient for the CO difference spectrum reading (the absorbance at 450 nm minus that at 490 nm) is 91 mM$^{-1}$cm$^{-1}$~\cite{Otey2003}, and we calculated the extinction coefficient at 398 nm for the 4-nitrophenolate group in our buffer to be 12,000 M$^{-1}$cm$^{-1}$.  Therefore, for 12-pNCA, the total number of substrate turnovers per P450 enzyme is 7.58 times the ratio of the 12-pNCA standard curve slope to the CO difference spectrum slope.  This indicates that our parent protein had about 250 12-pNCA turnovers per enzyme, compared to the 1,000 reported for a variant engineered for maximal 12-pNCA activity~\cite{Cirino2003}.  For the other substrates assayed with the 4-AAP assay, the extinction coefficient at 500 nm for the 4-AAP/phenol complex has been reported to be 4,800 ~\cite{Otey2004}.  However, we believe that this extinction coefficient could be of dubious accuracy for our data.  Depending on the exact type of phenolic compound created by P450 hydroxylation, the extinction coefficient for the 4-AAP/phenol complex may vary.  Assuming the extinction coefficient of 4,800 M$^{-1}$cm$^{-1}$ is accurate, then the total number of substrate turnovers per P450 enzyme is 19.0 times the ratio of the substrate standard curve slope to the CO difference spectrum slope.  Using this coefficient, the parent P450 had roughly 1,000 turnovers on 2-phenoxyethanol, 30 turnovers on propranolol, 400 turnovers on 11-phenoxyundecanoic acid, 50 turnovers on 2-amino-5-chlorobenzoxazole, and 80 turnovers on 1,2-methylenedioxybenzene.  The high activities on 2-phenoxyethanol and 11-phenoxyundecanoic acid are presumably due to the fact that lack of polar substituents on the aromatic ring allows these compounds to enter the hydrophobic P450 BM3 binding pocket~\cite{Haines2001} more easily than 12-pNCA.  However, we emphasize that the exact numerical values for the turnovers for these five substrates are questionable.  Definitive determination of the extinction coefficients would require analytical analysis of the enzymatic products for each P450 variant on each substrate, which is beyond the scope of this study.\pb

\subsection*{Analysis of activity data}
The raw activity values computed for the P450 variants are listed in Additional file \ref{add:activity_data}.  To analyze and display this data, we computed the fold change in activity of each variant relative to the R1-11 parent P450.  The fold change is simply the variant activity divided by the parent activity on each substrate, with the standard errors propagated to give an error on the fold change.  In Figures \ref{fig:heatmap} and \ref{fig:foldchanges}, these fold changes are displayed on a logarithmic scale so that each unit corresponds to a two-fold increase or decrease in activity.  In Figure \ref{fig:heatmap}, the substrates and the P450 variants have both been clustered, as shown by dendrograms on the side of the heat map.  The clustering was performed using the standard hierarchical clustering function of the R statistical package.  This is complete linkage hierarchical clustering, with the distances computed as the Euclidian distance between the logarithms of the fold changes in activity.  The standard errors on the fold changes in activity are not incorporated into Figure \ref{fig:heatmap} or any of the related analysis.  However, these standard errors are shown in Figure \ref{fig:foldchanges}; it is apparent from this figure that the errors tend to be much less than the fold changes in activity themselves.\pb

In Figure \ref{fig:changedistribution}, the histogram bins are logarithmically spaced so that each bin contains a $2^{0.5}$-fold range of activities.  For example, the histogram bin centered at one contains all variants with between $2^{-0.25} = 0.84$ and $2^{0.25} = 1.19$ fold the parental activity, while the bin centered at 1.5 contains all variants with between $2^{0.25} = 1.19$ and $2^{0.75} = 1.68$ fold the parental activity.\pb

The principal component analysis shown in Table \ref{tab:pca} was performed using the R statistical package, with inputs being the logarithms of the fold changes in activity.  Since these log fold changes in activity contained no arbitrary units (they were already normalized to the parent), the data was neither scaled nor zeroed before performing the analysis.  Table \ref{tab:pca} shows the composition and the percent of variance explained (the eigenvalue for that component divided by the sum of all eigenvalues) for the first two components.  The remaining four components were relatively unimportant, explaining 7\%, 5\%, 4\%, and 2\% of the total variance.\pb

\subsection*{Phylogenetic tree}
The phylogenetic tree shown in Figure \ref{fig:tree} is based on the number of nonsynonymous mutations the P450 variants have relative to the R-11 neutral evolution parent, as reported in \cite{Bloom2007c}.  Each of the P450s that evolved in a monomorphic population (prefix of M) are known to have diverged independently, and so are drawn on their own branch regardless of any sequence identity to other variants.  The exact phylogenetic relationship of the P450s that evolved in the polymorphic population (prefix of P) is not known, so they portion of the tree for these mutants was reconstructed by maximum parsimony.  The tree is based only on the nonsynonymous mutations, and all mutations weighted equally.  Full nucleotide and amino acid sequences of the P450s can be found in \cite{Bloom2007c}.
    
\section*{Authors contributions}
JDB, PR, and FHA designed the study.  JDB, PR, and ZL performed the experiments.  JDB and PR analyzed the data.  JDB and FHA wrote the paper.

\section*{Acknowledgements}
  \ifthenelse{\boolean{publ}}{\small}{}
 We thank Andrew Sawayama and Sabine Bastian for helpful comments.  JDB was supported by a Howard Hughes Medical Institute predoctoral fellowship.  ZL was supported by a summer undergraduate research fellowship from the California Institute of Technology.


{\ifthenelse{\boolean{publ}}{\footnotesize}{\small}
 \bibliographystyle{bmc_article}  
 \bibliography{/Users/bloom/references/references} }     


\begin{thebibliography}{10}
\providecommand{\url}[1]{[#1]}
\providecommand{\urlprefix}{}

\bibitem{O'Brien1999}
O'Brien PJ, Herschlag D: \textbf{Catalytic promiscuity and the evolution of new
  enzymatic activities}. \emph{Chemistry and Biology} 1999,
  \textbf{6}:R91--R105.

\bibitem{Aharoni2005}
Aharoni A, Gaidukov L, Khersonsky O, Gould SM, Roodveldt C, Tawfik DS:
  \textbf{The `evolvability' of promiscuous protein functions}. \emph{Nat.
  Genetics} 2005, \textbf{37}:73--76.

\bibitem{Kondrashov2005}
Kondrashov FA: \textbf{In search of the limits of evolution}. \emph{Nat.
  Genetics} 2005, \textbf{37}:9--10.

\bibitem{Chothia2003}
Chothia C, Gough J, Vogel C, Teichmann SA: \textbf{Evolution of the protein
  repertoire}. \emph{Science} 2003, \textbf{300}:1701--1703.

\bibitem{Copley2004}
Copley SD, Novak WRP, Babbitt PC: \textbf{Divergence of function in the
  thioredoxin fold suprafamily: evidence for evolution of peroxiredoxins from a
  thioredoxin-like ancestor}. \emph{Biochemistry} 2004,
  \textbf{43}:13981--13995.

\bibitem{Bridgham2006}
Bridgham JT, Carroll SM, Thornton JW: \textbf{Evolution of hormone-receptor
  complexity by molecular exploitation}. \emph{Science} 2006,
  \textbf{312}:97--101.

\bibitem{Copley2003}
Copley SD: \textbf{Enzymes with extra talents: moonlighting functions and
  catalytic promiscuity}. \emph{Curr. Opin. Chem. Biol.} 2003,
  \textbf{7}:265--272.

\bibitem{Khersonsky2006}
Khersonsky O, Roodveldt C, Tawfik DS: \textbf{Enzyme promiscuity: evolutionary
  and mechanistic aspects}. \emph{Curr. Opin. Chem. Biol.} 2006,
  \textbf{10}:498--508.

\bibitem{O'Brien2006}
O'Brien PJ: \textbf{Catalytic promiscuity and the divergent evolution of {DNA}
  repair enzymes}. \emph{Chem. Rev.} 2006, \textbf{106}:720--752.

\bibitem{O'Loughlin2007}
O'Loughlin TL, Greene DN, Matsumura I: \textbf{Diversification and
  specialization of {HIV} protease function during in vitro evolution}.
  \emph{Mol. Biol. Evol.} 2007, \textbf{23}:764--772.

\bibitem{Roodveldt2005}
Roodveldt C, Tawfik DS: \textbf{Shared promiscuous activities and evolutionary
  features in various members of the amidohydrolase superfamily}.
  \emph{Biochemistry} 2005, \textbf{44}:12728--12736.

\bibitem{Afriat2006}
Afriat L, Roodveldt C, Manco G, Tawfik DS: \textbf{The latent promiscuity of
  newly identified microbial lactonases is linked to a recently diverged
  phosphotriesterase}. \emph{Biochemistry} 2006, \textbf{45}:13677--13686.

\bibitem{Zuckerkandl1965}
Zuckerkandl E, Pauling L: \textbf{Evolutionary divergence and convergence in
  proteins}. In \emph{Evolving genes and proteins}, New York, NY: Academic
  Press 1965:97--166.

\bibitem{Kimura1983}
Kimura M: \emph{The Neutral Theory of Molecular Evolution}. Cambridge, U.K.:
  Cambridge Univ. Press 1983.

\bibitem{Blundell1975}
Blundell TL, Wood SP: \textbf{Is the evolution of insulin {Darwinian} or due to
  selectively neutral mutations?} \emph{Nature} 1975, \textbf{257}:197--203.

\bibitem{Bloom2007c}
Bloom JD, Lu Z, Chen D, Raval A, Venturelli OS, Arnold FH: \textbf{Neutral
  evolutions favors mutational robustness in sufficiently large populations}.
  \emph{Submitted} 2007. [Currently available on ArXiv preprint server at
  www.arxiv.org as arXiv:0704.1885v1].

\bibitem{Montellano1995}
de~Montellano PRO: \emph{Cytochrome P450: Structure, Mechanism, and
  Biochemistry}. New York: Plenum Press 1995.

\bibitem{Lewis2001}
Lewis DFV: \emph{Cytochromes {P450}: Structure and Function}. London: Taylor
  and Francis 2001.

\bibitem{Munro2002}
Munro AW, Leys DG, McLean KJ, Marshall KR, Ost TWB, Daff S, Miles CS, Chapman
  SK, Lysek DA, Moser CC, Page CC, Dutton PL: \textbf{P450 {BM}3: the very
  model of a modern flavocytochrome}. \emph{Trends Biochem. Sci.} 2002,
  \textbf{27}:250--257.

\bibitem{DePristo2005}
DePristo MA, Weinreich DM, Hartl DL: \textbf{Missense meanderings in sequence
  space: a biophysical view of protein evolution}. \emph{Nat. Rev. Genetics}
  2005, \textbf{6}:678--687.

\bibitem{Bloom2007}
Bloom JD, Raval A, Wilke CO: \textbf{Thermodynamics of neutral protein
  evolution}. \emph{Genetics} 2007, \textbf{175}:255--266.

\bibitem{Bloom2005}
Bloom JD, Silberg JJ, Wilke CO, Drummond DA, Adami C, Arnold FH:
  \textbf{Thermodynamic prediction of protein neutrality}. \emph{Proc. Natl.
  Acad. Sci. USA} 2005, \textbf{102}:606--611.

\bibitem{Besenmatter2007}
Besenmatter W, Kast P, Hilvert D: \textbf{Relative tolerance of mesostable and
  thermostable protein homologs to extensive mutation}. \emph{Proteins} 2007,
  \textbf{66}:500--506.

\bibitem{Bloom2006}
Bloom JD, Labthavikul ST, Otey CR, Arnold FH: \textbf{Protein stability
  promotes evolvability}. \emph{Proc. Natl. Acad. Sci. USA} 2006,
  \textbf{103}:5869--5874.

\bibitem{Otey2005}
Otey CR, Bandara G, Lalonde J, Takahashi K, Arnold FH: \textbf{Preparation of
  human metabolites of propranolol using laboratory-evolved bacterial
  cytochromes {P450}}. \emph{Biotechnol. and Bioeng.} 2005,
  \textbf{93}:494--499.

\bibitem{Lasker1982}
Lasker JM, Huang MT, Conney AH: \textbf{In vitro activation of zoxazolamine
  metabolism by flavone}. \emph{Science} 1982, \textbf{216}:1419--1421.

\bibitem{Cirino2003}
Cirino PC, Arnold FH: \textbf{A self-sufficient peroxide-driven hydroxylation
  biocatalyst}. \emph{Angew. Chem. Int. Ed.} 2003, \textbf{42}:3299--3301.

\bibitem{Taverna2002}
Taverna DM, Goldstein RA: \textbf{Why are proteins marginally stable?}
  \emph{Proteins} 2002, \textbf{46}:105--109.

\bibitem{Haines2001}
Haines DC, Tomchick DR, Machius M, Peterson JA: \textbf{Pivotal role of water
  in the mechanism of {P450BM-3}}. \emph{Biochemistry} 2001,
  \textbf{40}:13456--13465.

\bibitem{Li2001}
Li QS, Ogawa J, Schmid RD, Shimizu S: \textbf{Engineering cytochrome {P450
  BM-3} for oxidation of polycyclic aromatic hydrocarbons}. \emph{Appl.
  Environ. Microbiol.} 2001, \textbf{67}:5735--5739.

\bibitem{Li2001b}
Li QS, Schwaneberg U, Fischer M, Schmitt J, Pleiss J, Lutz-Wahl S, Schmid RD:
  \textbf{Rational evolution of a medium chain-specific cytochrome {P-450 BM-3}
  variant}. \emph{Biochim. Biophys. Acta} 2001, \textbf{1545}:114--121.

\bibitem{Pylypenko2004}
Pylypenko O, Schlicting I: \textbf{Structural aspects of ligand binding to and
  electron transfer in bacterial and fungal {P450s}}. \emph{Annu. Rev.
  Biochem.} 2004, \textbf{73}:991--1018.

\bibitem{Modi1996}
Modi S, Sutcliffe MJ, Primrose WU, Lian LY, Roberts GCK: \textbf{The catalytic
  mechanism of cytochrome {P450 BM3} involves a 6 angstrom movement of the
  bound substrate on reduction}. \emph{Nat. Struct. Biol.} 1996,
  \textbf{3}:414--417.

\bibitem{Glieder2002}
Glieder A, Farinas ET, Arnold FH: \textbf{Laboratory evolution of a soluble,
  self-sufficient, highly active alkane hydroxylase}. \emph{Nat. Biotech.}
  2002, \textbf{20}:1135--1139.

\bibitem{Meinhold2005}
Meinhold P, Peters MW, Chen MMY, Takahashi K, Arnold FH: \textbf{Direct
  conversion of ethane to ethanol by engineered cytochrome {P450 BM3}}.
  \emph{ChemBioChem} 2005, \textbf{6}:1765--1768.

\bibitem{Loida1993}
Loida PJ, Sligar SG: \textbf{Molecular recognition in cytochrome {P450} -
  mechanism for the control of uncoupling reactions}. \emph{Biochemistry} 1993,
  \textbf{32}:11530--11538.

\bibitem{Bernhardt2006}
Bernhardt R: \textbf{Cytochromes {P450} as versatile biocatalysts}. \emph{J.
  Biotechnol.} 2006, \textbf{124}:128--145.

\bibitem{Amitai2007}
Amitai G, Gupta RD, Tawfik DS: \textbf{Latent evolutionary potentials under the
  neutral mutation drift of an enzyme}. \emph{HFSP Journal} 2007, \textbf{in
  press}.

\bibitem{Landwehr2007}
Landwehr M, Carbone M, Otey CR, Li Y, Arnold FH: \textbf{Diversification of
  catalytic function in a synthetic family of chimeric cytochrome {P450s}}.
  \emph{Chem. Biol.} 2007, \textbf{14}:269--278.

\bibitem{Varadarajan2005}
Varadarajan N, Gam J, Olsen MJ, Georgiou G, Iverson BL: \textbf{Engineering of
  protease variants exhibiting high catalytic activity and exquisite substrate
  selectivity}. \emph{Proc. Natl. Acad. Sci. USA} 2005,
  \textbf{102}:6855--6860.

\bibitem{Huynen1996}
Huynen MA, Stadler PF, Fontana W: \textbf{Smoothness within ruggedness: the
  role of neutrality in adaptation}. \emph{Proc. Natl. Acad. Sci. USA} 1996,
  \textbf{93}:397--401.

\bibitem{Huynen1996b}
Huynen MA: \textbf{Exploring phenotype space through neutral evolution}.
  \emph{J. Mol. Evol.} 1996, \textbf{43}:165--169.

\bibitem{Fontana1998}
Fontana W, Schuster P: \textbf{Continuity in evolution: on the nature of
  transitions}. \emph{Science} 1998, \textbf{280}:1451--1455.

\bibitem{vanNimwegen1997}
van Nimwegen E, Crutchfield JP, Mitchell M: \textbf{Finite populations induce
  metastability in evolutionary search}. \emph{Physics Letters A} 1997,
  \textbf{229}:144--150.

\bibitem{vanNimwegen2000}
van Nimwegen E, Crutchfield JP: \textbf{Metastable evolutionary dynamics:
  crossing fitness barriers or escaping via neutral paths?} \emph{Bull. Math.
  Biol.} 2000, \textbf{62}:799--848.

\bibitem{Serrano1993}
Serrano L, Day AG, Fersht AR: \textbf{Step-wise mutation of barnase to binase:
  a procedure for engineering increased stability of proteins and an
  experimental analysis of the evolution of protein stability}. \emph{J. Mol.
  Biol.} 1993, \textbf{233}:305--312.

\bibitem{Barnes1991}
Barnes HJ, Arlotto MP, Waterman MR: \textbf{Expression and enzymatic activity
  of recombinant cytochrome {P450} 17 $\alpha$-hydroxylase in
  \textit{Escherichia coli}}. \emph{Proc. Natl. Acad. Sci. USA} 1991,
  \textbf{88}:5597--5601.

\bibitem{Otey2003}
Otey CR: \textbf{High-throughput carbon monoxide binding assay for cytochromes
  {P}450}. In \emph{Directed enzyme evolution: screening and selection
  methods}, \emph{Volume 230 of \emph{Methods in Molecular Biology}}. Edited by
  Arnold FH, Georgiou G, Humana press 2003:137--139.

\bibitem{Schwaneberg1999}
Schwaneberg U, Schmidt-Dannert C, Schmitt J, Schmid RD: \textbf{A continuous
  spectrophotometric assay for {P450 BM-3}, a fatty acid hydroxylating enzyme,
  and its mutant {F87A}}. \emph{Analytical Biochemistry} 1999,
  \textbf{269}:359--366.

\bibitem{Otey2003b}
Otey CR, Joern JM: \textbf{High-throughput screen for aromatic hydroxylation}.
  In \emph{Directed enzyme evolution: screening and selection methods},
  \emph{Volume 230 of \emph{Methods in Molecular Biology}}. Edited by Arnold
  FH, Georgiou G, Humana press 2003:141--148.

\bibitem{Otey2004}
Otey CR, Silberg JJ, Voigt CA, Endelman JB, Bandara G, Arnold FH:
  \textbf{Functional evolution and structural conservation in chimeric
  cytochromes P450: calibrating a structure-guided approach}. \emph{Chem.
  Biol.} 2004, \textbf{11}:309--318.

\end{thebibliography}

\newcommand{\BMCxmlcomment}[1]{}

\BMCxmlcomment{

<refgrp>

<bibl id="B1">
  <title><p>Catalytic promiscuity and the evolution of new enzymatic
  activities</p></title>
  <aug>
    <au><snm>O'Brien</snm><fnm>PJ</fnm></au>
    <au><snm>Herschlag</snm><fnm>D</fnm></au>
  </aug>
  <source>Chemistry and Biology</source>
  <pubdate>1999</pubdate>
  <volume>6</volume>
  <fpage>R91</fpage>
  <lpage>R105</lpage>
</bibl>

<bibl id="B2">
  <title><p>The `evolvability' of promiscuous protein functions</p></title>
  <aug>
    <au><snm>Aharoni</snm><fnm>A</fnm></au>
    <au><snm>Gaidukov</snm><fnm>L</fnm></au>
    <au><snm>Khersonsky</snm><fnm>O</fnm></au>
    <au><snm>Gould</snm><fnm>SM</fnm></au>
    <au><snm>Roodveldt</snm><fnm>C</fnm></au>
    <au><snm>Tawfik</snm><fnm>DS</fnm></au>
  </aug>
  <source>Nat. Genetics</source>
  <pubdate>2005</pubdate>
  <volume>37</volume>
  <fpage>73</fpage>
  <lpage>76</lpage>
</bibl>

<bibl id="B3">
  <title><p>In search of the limits of evolution</p></title>
  <aug>
    <au><snm>Kondrashov</snm><fnm>FA</fnm></au>
  </aug>
  <source>Nat. Genetics</source>
  <pubdate>2005</pubdate>
  <volume>37</volume>
  <fpage>9</fpage>
  <lpage>10</lpage>
</bibl>

<bibl id="B4">
  <title><p>Evolution of the protein repertoire</p></title>
  <aug>
    <au><snm>Chothia</snm><fnm>C.</fnm></au>
    <au><snm>Gough</snm><fnm>J.</fnm></au>
    <au><snm>Vogel</snm><fnm>C.</fnm></au>
    <au><snm>Teichmann</snm><fnm>S. A</fnm></au>
  </aug>
  <source>Science</source>
  <pubdate>2003</pubdate>
  <volume>300</volume>
  <fpage>1701</fpage>
  <lpage>1703</lpage>
</bibl>

<bibl id="B5">
  <title><p>Divergence of function in the thioredoxin fold suprafamily:
  evidence for evolution of peroxiredoxins from a thioredoxin-like
  ancestor</p></title>
  <aug>
    <au><snm>Copley</snm><fnm>S. D.</fnm></au>
    <au><snm>Novak</snm><fnm>W. R. P.</fnm></au>
    <au><snm>Babbitt</snm><fnm>P. C.</fnm></au>
  </aug>
  <source>Biochemistry</source>
  <pubdate>2004</pubdate>
  <volume>43</volume>
  <fpage>13981</fpage>
  <lpage>13995</lpage>
</bibl>

<bibl id="B6">
  <title><p>Evolution of hormone-receptor complexity by molecular
  exploitation</p></title>
  <aug>
    <au><snm>Bridgham</snm><fnm>JT</fnm></au>
    <au><snm>Carroll</snm><fnm>SM</fnm></au>
    <au><snm>Thornton</snm><fnm>JW</fnm></au>
  </aug>
  <source>Science</source>
  <pubdate>2006</pubdate>
  <volume>312</volume>
  <fpage>97</fpage>
  <lpage>101</lpage>
</bibl>

<bibl id="B7">
  <title><p>Enzymes with extra talents: moonlighting functions and catalytic
  promiscuity</p></title>
  <aug>
    <au><snm>Copley</snm><fnm>SD</fnm></au>
  </aug>
  <source>Curr. Opin. Chem. Biol.</source>
  <pubdate>2003</pubdate>
  <volume>7</volume>
  <fpage>265</fpage>
  <lpage>272</lpage>
</bibl>

<bibl id="B8">
  <title><p>Enzyme promiscuity: evolutionary and mechanistic
  aspects</p></title>
  <aug>
    <au><snm>Khersonsky</snm><fnm>O</fnm></au>
    <au><snm>Roodveldt</snm><fnm>C</fnm></au>
    <au><snm>Tawfik</snm><fnm>DS</fnm></au>
  </aug>
  <source>Curr. Opin. Chem. Biol.</source>
  <pubdate>2006</pubdate>
  <volume>10</volume>
  <fpage>498</fpage>
  <lpage>508</lpage>
</bibl>

<bibl id="B9">
  <title><p>Catalytic promiscuity and the divergent evolution of {DNA} repair
  enzymes</p></title>
  <aug>
    <au><snm>O'Brien</snm><fnm>PJ</fnm></au>
  </aug>
  <source>Chem. Rev.</source>
  <pubdate>2006</pubdate>
  <volume>106</volume>
  <fpage>720</fpage>
  <lpage>-752</lpage>
</bibl>

<bibl id="B10">
  <title><p>Diversification and specialization of {HIV} protease function
  during in vitro evolution</p></title>
  <aug>
    <au><snm>O'Loughlin</snm><fnm>TL</fnm></au>
    <au><snm>Greene</snm><fnm>DN</fnm></au>
    <au><snm>Matsumura</snm><fnm>I</fnm></au>
  </aug>
  <source>Mol. Biol. Evol.</source>
  <pubdate>2007</pubdate>
  <volume>23</volume>
  <fpage>764</fpage>
  <lpage>772</lpage>
</bibl>

<bibl id="B11">
  <title><p>Shared promiscuous activities and evolutionary features in various
  members of the amidohydrolase superfamily</p></title>
  <aug>
    <au><snm>Roodveldt</snm><fnm>C</fnm></au>
    <au><snm>Tawfik</snm><fnm>DS</fnm></au>
  </aug>
  <source>Biochemistry</source>
  <pubdate>2005</pubdate>
  <volume>44</volume>
  <fpage>12728</fpage>
  <lpage>12736</lpage>
</bibl>

<bibl id="B12">
  <title><p>The latent promiscuity of newly identified microbial lactonases is
  linked to a recently diverged phosphotriesterase</p></title>
  <aug>
    <au><snm>Afriat</snm><fnm>L</fnm></au>
    <au><snm>Roodveldt</snm><fnm>C</fnm></au>
    <au><snm>Manco</snm><fnm>G</fnm></au>
    <au><snm>Tawfik</snm><fnm>DS</fnm></au>
  </aug>
  <source>Biochemistry</source>
  <pubdate>2006</pubdate>
  <volume>45</volume>
  <fpage>13677</fpage>
  <lpage>13686</lpage>
</bibl>

<bibl id="B13">
  <title><p>Evolutionary divergence and convergence in proteins</p></title>
  <aug>
    <au><snm>Zuckerkandl</snm><fnm>E.</fnm></au>
    <au><snm>Pauling</snm><fnm>L.</fnm></au>
  </aug>
  <source>Evolving genes and proteins</source>
  <publisher>New York, NY: Academic Press</publisher>
  <pubdate>1965</pubdate>
  <fpage>97</fpage>
  <lpage>-166</lpage>
</bibl>

<bibl id="B14">
  <title><p>The Neutral Theory of Molecular Evolution</p></title>
  <aug>
    <au><snm>Kimura</snm><fnm>M.</fnm></au>
  </aug>
  <publisher>Cambridge, U.K.: Cambridge Univ. Press</publisher>
  <pubdate>1983</pubdate>
</bibl>

<bibl id="B15">
  <title><p>Is the evolution of insulin {Darwinian} or due to selectively
  neutral mutations?</p></title>
  <aug>
    <au><snm>Blundell</snm><fnm>T. L.</fnm></au>
    <au><snm>Wood</snm><fnm>S. P.</fnm></au>
  </aug>
  <source>Nature</source>
  <pubdate>1975</pubdate>
  <volume>257</volume>
  <fpage>197</fpage>
  <lpage>203</lpage>
</bibl>

<bibl id="B16">
  <title><p>Neutral evolutions favors mutational robustness in sufficiently
  large populations</p></title>
  <aug>
    <au><snm>Bloom</snm><fnm>JD</fnm></au>
    <au><snm>Lu</snm><fnm>Z</fnm></au>
    <au><snm>Chen</snm><fnm>D</fnm></au>
    <au><snm>Raval</snm><fnm>A</fnm></au>
    <au><snm>Venturelli</snm><fnm>OS</fnm></au>
    <au><snm>Arnold</snm><fnm>FH</fnm></au>
  </aug>
  <source>Submitted</source>
  <pubdate>2007</pubdate>
  <note>Currently available on ArXiv preprint server at www.arxiv.org as
  arXiv:0704.1885v1</note>
</bibl>

<bibl id="B17">
  <title><p>Cytochrome P450: Structure, Mechanism, and Biochemistry</p></title>
  <aug>
    <au><snm>Montellano</snm><fnm>PRO</fnm></au>
  </aug>
  <publisher>New York: Plenum Press</publisher>
  <pubdate>1995</pubdate>
</bibl>

<bibl id="B18">
  <title><p>Cytochromes {P450}: Structure and Function</p></title>
  <aug>
    <au><snm>Lewis</snm><fnm>D. F. V.</fnm></au>
  </aug>
  <publisher>London: Taylor and Francis</publisher>
  <pubdate>2001</pubdate>
</bibl>

<bibl id="B19">
  <title><p>P450 {BM}3: the very model of a modern flavocytochrome</p></title>
  <aug>
    <au><snm>Munro</snm><fnm>A. W.</fnm></au>
    <au><snm>Leys</snm><fnm>D. G.</fnm></au>
    <au><snm>McLean</snm><fnm>K. J.</fnm></au>
    <au><snm>Marshall</snm><fnm>K. R.</fnm></au>
    <au><snm>Ost</snm><fnm>T. W. B.</fnm></au>
    <au><snm>Daff</snm><fnm>S.</fnm></au>
    <au><snm>Miles</snm><fnm>C. S.</fnm></au>
    <au><snm>Chapman</snm><fnm>S. K.</fnm></au>
    <au><snm>Lysek</snm><fnm>D. A.</fnm></au>
    <au><snm>Moser</snm><fnm>C. C.</fnm></au>
    <au><snm>Page</snm><fnm>C. C.</fnm></au>
    <au><snm>Dutton</snm><fnm>P. L.</fnm></au>
  </aug>
  <source>Trends Biochem. Sci.</source>
  <pubdate>2002</pubdate>
  <volume>27</volume>
  <fpage>250</fpage>
  <lpage>-257</lpage>
</bibl>

<bibl id="B20">
  <title><p>Missense meanderings in sequence space: a biophysical view of
  protein evolution</p></title>
  <aug>
    <au><snm>DePristo</snm><fnm>M. A.</fnm></au>
    <au><snm>Weinreich</snm><fnm>D. M.</fnm></au>
    <au><snm>Hartl</snm><fnm>D. L.</fnm></au>
  </aug>
  <source>Nat. Rev. Genetics</source>
  <pubdate>2005</pubdate>
  <volume>6</volume>
  <fpage>678</fpage>
  <lpage>-687</lpage>
</bibl>

<bibl id="B21">
  <title><p>Thermodynamics of neutral protein evolution</p></title>
  <aug>
    <au><snm>Bloom</snm><fnm>J. D.</fnm></au>
    <au><snm>Raval</snm><fnm>A.</fnm></au>
    <au><snm>Wilke</snm><fnm>C. O.</fnm></au>
  </aug>
  <source>Genetics</source>
  <pubdate>2007</pubdate>
  <volume>175</volume>
  <fpage>255</fpage>
  <lpage>266</lpage>
</bibl>

<bibl id="B22">
  <title><p>Thermodynamic prediction of protein neutrality</p></title>
  <aug>
    <au><snm>Bloom</snm><fnm>J. D.</fnm></au>
    <au><snm>Silberg</snm><fnm>J. J.</fnm></au>
    <au><snm>Wilke</snm><fnm>C. O.</fnm></au>
    <au><snm>Drummond</snm><fnm>D. A.</fnm></au>
    <au><snm>Adami</snm><fnm>C.</fnm></au>
    <au><snm>Arnold</snm><fnm>F. H.</fnm></au>
  </aug>
  <source>Proc. Natl. Acad. Sci. USA</source>
  <pubdate>2005</pubdate>
  <volume>102</volume>
  <fpage>606</fpage>
  <lpage>-611</lpage>
</bibl>

<bibl id="B23">
  <title><p>Relative tolerance of mesostable and thermostable protein homologs
  to extensive mutation</p></title>
  <aug>
    <au><snm>Besenmatter</snm><fnm>W.</fnm></au>
    <au><snm>Kast</snm><fnm>P.</fnm></au>
    <au><snm>Hilvert</snm><fnm>D.</fnm></au>
  </aug>
  <source>Proteins</source>
  <pubdate>2007</pubdate>
  <volume>66</volume>
  <fpage>500</fpage>
  <lpage>506</lpage>
</bibl>

<bibl id="B24">
  <title><p>Protein stability promotes evolvability</p></title>
  <aug>
    <au><snm>Bloom</snm><fnm>J. D.</fnm></au>
    <au><snm>Labthavikul</snm><fnm>S. T.</fnm></au>
    <au><snm>Otey</snm><fnm>C. R.</fnm></au>
    <au><snm>Arnold</snm><fnm>F. H.</fnm></au>
  </aug>
  <source>Proc. Natl. Acad. Sci. USA</source>
  <pubdate>2006</pubdate>
  <volume>103</volume>
  <fpage>5869</fpage>
  <lpage>-5874</lpage>
</bibl>

<bibl id="B25">
  <title><p>Preparation of human metabolites of propranolol using
  laboratory-evolved bacterial cytochromes {P450}</p></title>
  <aug>
    <au><snm>Otey</snm><fnm>C. R.</fnm></au>
    <au><snm>Bandara</snm><fnm>G.</fnm></au>
    <au><snm>Lalonde</snm><fnm>J.</fnm></au>
    <au><snm>Takahashi</snm><fnm>K.</fnm></au>
    <au><snm>Arnold</snm><fnm>F. H.</fnm></au>
  </aug>
  <source>Biotechnol. and Bioeng.</source>
  <pubdate>2005</pubdate>
  <volume>93</volume>
  <fpage>494</fpage>
  <lpage>499</lpage>
</bibl>

<bibl id="B26">
  <title><p>In vitro activation of zoxazolamine metabolism by
  flavone</p></title>
  <aug>
    <au><snm>Lasker</snm><fnm>J. M.</fnm></au>
    <au><snm>Huang</snm><fnm>M. T.</fnm></au>
    <au><snm>Conney</snm><fnm>A. H.</fnm></au>
  </aug>
  <source>Science</source>
  <pubdate>1982</pubdate>
  <volume>216</volume>
  <fpage>1419</fpage>
  <lpage>1421</lpage>
</bibl>

<bibl id="B27">
  <title><p>A self-sufficient peroxide-driven hydroxylation
  biocatalyst</p></title>
  <aug>
    <au><snm>Cirino</snm><fnm>P. C.</fnm></au>
    <au><snm>Arnold</snm><fnm>F. H.</fnm></au>
  </aug>
  <source>Angew. Chem. Int. Ed.</source>
  <pubdate>2003</pubdate>
  <volume>42</volume>
  <fpage>3299</fpage>
  <lpage>-3301</lpage>
</bibl>

<bibl id="B28">
  <title><p>Why are proteins marginally stable?</p></title>
  <aug>
    <au><snm>Taverna</snm><fnm>D. M.</fnm></au>
    <au><snm>Goldstein</snm><fnm>R. A.</fnm></au>
  </aug>
  <source>Proteins</source>
  <pubdate>2002</pubdate>
  <volume>46</volume>
  <fpage>105</fpage>
  <lpage>-109</lpage>
</bibl>

<bibl id="B29">
  <title><p>Pivotal role of water in the mechanism of {P450BM-3}</p></title>
  <aug>
    <au><snm>Haines</snm><fnm>D. C.</fnm></au>
    <au><snm>Tomchick</snm><fnm>D. R.</fnm></au>
    <au><snm>Machius</snm><fnm>M.</fnm></au>
    <au><snm>Peterson</snm><fnm>J. A.</fnm></au>
  </aug>
  <source>Biochemistry</source>
  <pubdate>2001</pubdate>
  <volume>40</volume>
  <fpage>13456</fpage>
  <lpage>-13465</lpage>
</bibl>

<bibl id="B30">
  <title><p>Engineering cytochrome {P450 BM-3} for oxidation of polycyclic
  aromatic hydrocarbons</p></title>
  <aug>
    <au><snm>Li</snm><fnm>Q. S.</fnm></au>
    <au><snm>Ogawa</snm><fnm>J.</fnm></au>
    <au><snm>Schmid</snm><fnm>R. D.</fnm></au>
    <au><snm>Shimizu</snm><fnm>S.</fnm></au>
  </aug>
  <source>Appl. Environ. Microbiol.</source>
  <pubdate>2001</pubdate>
  <volume>67</volume>
  <fpage>5735</fpage>
  <lpage>5739</lpage>
</bibl>

<bibl id="B31">
  <title><p>Rational evolution of a medium chain-specific cytochrome {P-450
  BM-3} variant</p></title>
  <aug>
    <au><snm>Li</snm><fnm>Q. S.</fnm></au>
    <au><snm>Schwaneberg</snm><fnm>U.</fnm></au>
    <au><snm>Fischer</snm><fnm>M.</fnm></au>
    <au><snm>Schmitt</snm><fnm>J.</fnm></au>
    <au><snm>Pleiss</snm><fnm>J.</fnm></au>
    <au><snm>Lutz Wahl</snm><fnm>S.</fnm></au>
    <au><snm>Schmid</snm><fnm>R. D.</fnm></au>
  </aug>
  <source>Biochim. Biophys. Acta</source>
  <pubdate>2001</pubdate>
  <volume>1545</volume>
  <fpage>114</fpage>
  <lpage>121</lpage>
</bibl>

<bibl id="B32">
  <title><p>Structural aspects of ligand binding to and electron transfer in
  bacterial and fungal {P450s}</p></title>
  <aug>
    <au><snm>Pylypenko</snm><fnm>O.</fnm></au>
    <au><snm>Schlicting</snm><fnm>I.</fnm></au>
  </aug>
  <source>Annu. Rev. Biochem.</source>
  <pubdate>2004</pubdate>
  <volume>73</volume>
  <fpage>991</fpage>
  <lpage>1018</lpage>
</bibl>

<bibl id="B33">
  <title><p>The catalytic mechanism of cytochrome {P450 BM3} involves a 6
  angstrom movement of the bound substrate on reduction</p></title>
  <aug>
    <au><snm>Modi</snm><fnm>S.</fnm></au>
    <au><snm>Sutcliffe</snm><fnm>M. J.</fnm></au>
    <au><snm>Primrose</snm><fnm>W. U.</fnm></au>
    <au><snm>Lian</snm><fnm>L. Y.</fnm></au>
    <au><snm>Roberts</snm><fnm>G. C. K.</fnm></au>
  </aug>
  <source>Nat. Struct. Biol.</source>
  <pubdate>1996</pubdate>
  <volume>3</volume>
  <fpage>414</fpage>
  <lpage>417</lpage>
</bibl>

<bibl id="B34">
  <title><p>Laboratory evolution of a soluble, self-sufficient, highly active
  alkane hydroxylase</p></title>
  <aug>
    <au><snm>Glieder</snm><fnm>A.</fnm></au>
    <au><snm>Farinas</snm><fnm>E. T.</fnm></au>
    <au><snm>Arnold</snm><fnm>F. H.</fnm></au>
  </aug>
  <source>Nat. Biotech.</source>
  <pubdate>2002</pubdate>
  <volume>20</volume>
  <fpage>1135</fpage>
  <lpage>1139</lpage>
</bibl>

<bibl id="B35">
  <title><p>Direct conversion of ethane to ethanol by engineered cytochrome
  {P450 BM3}</p></title>
  <aug>
    <au><snm>Meinhold</snm><fnm>P.</fnm></au>
    <au><snm>Peters</snm><fnm>M. W.</fnm></au>
    <au><snm>Chen</snm><fnm>M. M. Y.</fnm></au>
    <au><snm>Takahashi</snm><fnm>K.</fnm></au>
    <au><snm>Arnold</snm><fnm>F. H.</fnm></au>
  </aug>
  <source>ChemBioChem</source>
  <pubdate>2005</pubdate>
  <volume>6</volume>
  <fpage>1765</fpage>
  <lpage>1768</lpage>
</bibl>

<bibl id="B36">
  <title><p>Molecular recognition in cytochrome {P450} - mechanism for the
  control of uncoupling reactions</p></title>
  <aug>
    <au><snm>Loida</snm><fnm>P. J.</fnm></au>
    <au><snm>Sligar</snm><fnm>S. G.</fnm></au>
  </aug>
  <source>Biochemistry</source>
  <pubdate>1993</pubdate>
  <volume>32</volume>
  <fpage>11530</fpage>
  <lpage>11538</lpage>
</bibl>

<bibl id="B37">
  <title><p>Cytochromes {P450} as versatile biocatalysts</p></title>
  <aug>
    <au><snm>Bernhardt</snm><fnm>R.</fnm></au>
  </aug>
  <source>J. Biotechnol.</source>
  <pubdate>2006</pubdate>
  <volume>124</volume>
  <fpage>128</fpage>
  <lpage>145</lpage>
</bibl>

<bibl id="B38">
  <title><p>Latent evolutionary potentials under the neutral mutation drift of
  an enzyme</p></title>
  <aug>
    <au><snm>Amitai</snm><fnm>G</fnm></au>
    <au><snm>Gupta</snm><fnm>RD</fnm></au>
    <au><snm>Tawfik</snm><fnm>DS</fnm></au>
  </aug>
  <source>HFSP Journal</source>
  <pubdate>2007</pubdate>
  <volume>in press</volume>
</bibl>

<bibl id="B39">
  <title><p>Diversification of catalytic function in a synthetic family of
  chimeric cytochrome {P450s}</p></title>
  <aug>
    <au><snm>Landwehr</snm><fnm>M.</fnm></au>
    <au><snm>Carbone</snm><fnm>M.</fnm></au>
    <au><snm>Otey</snm><fnm>C. R.</fnm></au>
    <au><snm>Li</snm><fnm>Y.</fnm></au>
    <au><snm>Arnold</snm><fnm>F. H.</fnm></au>
  </aug>
  <source>Chem. Biol.</source>
  <pubdate>2007</pubdate>
  <volume>14</volume>
  <fpage>269</fpage>
  <lpage>278</lpage>
</bibl>

<bibl id="B40">
  <title><p>Engineering of protease variants exhibiting high catalytic activity
  and exquisite substrate selectivity</p></title>
  <aug>
    <au><snm>Varadarajan</snm><fnm>N.</fnm></au>
    <au><snm>Gam</snm><fnm>J.</fnm></au>
    <au><snm>Olsen</snm><fnm>M. J.</fnm></au>
    <au><snm>Georgiou</snm><fnm>G.</fnm></au>
    <au><snm>Iverson</snm><fnm>B. L.</fnm></au>
  </aug>
  <source>Proc. Natl. Acad. Sci. USA</source>
  <pubdate>2005</pubdate>
  <volume>102</volume>
  <fpage>6855</fpage>
  <lpage>6860</lpage>
</bibl>

<bibl id="B41">
  <title><p>Smoothness within ruggedness: the role of neutrality in
  adaptation</p></title>
  <aug>
    <au><snm>Huynen</snm><fnm>M. A.</fnm></au>
    <au><snm>Stadler</snm><fnm>P. F.</fnm></au>
    <au><snm>Fontana</snm><fnm>W.</fnm></au>
  </aug>
  <source>Proc. Natl. Acad. Sci. USA</source>
  <pubdate>1996</pubdate>
  <volume>93</volume>
  <fpage>397</fpage>
  <lpage>-401</lpage>
</bibl>

<bibl id="B42">
  <title><p>Exploring phenotype space through neutral evolution</p></title>
  <aug>
    <au><snm>Huynen</snm><fnm>M. A.</fnm></au>
  </aug>
  <source>J. Mol. Evol.</source>
  <pubdate>1996</pubdate>
  <volume>43</volume>
  <fpage>165</fpage>
  <lpage>169</lpage>
</bibl>

<bibl id="B43">
  <title><p>Continuity in evolution: on the nature of transitions</p></title>
  <aug>
    <au><snm>Fontana</snm><fnm>W.</fnm></au>
    <au><snm>Schuster</snm><fnm>P.</fnm></au>
  </aug>
  <source>Science</source>
  <pubdate>1998</pubdate>
  <volume>280</volume>
  <fpage>1451</fpage>
  <lpage>1455</lpage>
</bibl>

<bibl id="B44">
  <title><p>Finite populations induce metastability in evolutionary
  search</p></title>
  <aug>
    <au><snm>Nimwegen</snm><fnm>E.</fnm></au>
    <au><snm>Crutchfield</snm><fnm>J. P.</fnm></au>
    <au><snm>Mitchell</snm><fnm>M.</fnm></au>
  </aug>
  <source>Physics Letters A</source>
  <pubdate>1997</pubdate>
  <volume>229</volume>
  <fpage>144</fpage>
  <lpage>150</lpage>
</bibl>

<bibl id="B45">
  <title><p>Metastable evolutionary dynamics: crossing fitness barriers or
  escaping via neutral paths?</p></title>
  <aug>
    <au><snm>Nimwegen</snm><fnm>E.</fnm></au>
    <au><snm>Crutchfield</snm><fnm>J. P.</fnm></au>
  </aug>
  <source>Bull. Math. Biol.</source>
  <pubdate>2000</pubdate>
  <volume>62</volume>
  <fpage>799</fpage>
  <lpage>848</lpage>
</bibl>

<bibl id="B46">
  <title><p>Step-wise mutation of barnase to binase: a procedure for
  engineering increased stability of proteins and an experimental analysis of
  the evolution of protein stability</p></title>
  <aug>
    <au><snm>Serrano</snm><fnm>L.</fnm></au>
    <au><snm>Day</snm><fnm>A. G.</fnm></au>
    <au><snm>Fersht</snm><fnm>A. R.</fnm></au>
  </aug>
  <source>J. Mol. Biol.</source>
  <pubdate>1993</pubdate>
  <volume>233</volume>
  <fpage>305</fpage>
  <lpage>-312</lpage>
</bibl>

<bibl id="B47">
  <title><p>Expression and enzymatic activity of recombinant cytochrome {P450}
  17 $\alpha$-hydroxylase in \textit{Escherichia coli}</p></title>
  <aug>
    <au><snm>Barnes</snm><fnm>H. J.</fnm></au>
    <au><snm>Arlotto</snm><fnm>M. P.</fnm></au>
    <au><snm>Waterman</snm><fnm>M. R.</fnm></au>
  </aug>
  <source>Proc. Natl. Acad. Sci. USA</source>
  <pubdate>1991</pubdate>
  <volume>88</volume>
  <fpage>5597</fpage>
  <lpage>5601</lpage>
</bibl>

<bibl id="B48">
  <title><p>High-throughput carbon monoxide binding assay for cytochromes
  {P}450</p></title>
  <aug>
    <au><snm>Otey</snm><fnm>C. R.</fnm></au>
  </aug>
  <source>Directed enzyme evolution: screening and selection methods</source>
  <publisher>Humana press</publisher>
  <editor>F. H. Arnold and G. Georgiou</editor>
  <series><title><p>Methods in Molecular Biology</p></title></series>
  <section><title><p>13</p></title></section>
  <pubdate>2003</pubdate>
  <volume>230</volume>
  <fpage>137</fpage>
  <lpage>-139</lpage>
</bibl>

<bibl id="B49">
  <title><p>A continuous spectrophotometric assay for {P450 BM-3}, a fatty acid
  hydroxylating enzyme, and its mutant {F87A}</p></title>
  <aug>
    <au><snm>Schwaneberg</snm><fnm>U</fnm></au>
    <au><snm>Schmidt Dannert</snm><fnm>C</fnm></au>
    <au><snm>Schmitt</snm><fnm>J</fnm></au>
    <au><snm>Schmid</snm><fnm>RD</fnm></au>
  </aug>
  <source>Analytical Biochemistry</source>
  <pubdate>1999</pubdate>
  <volume>269</volume>
  <fpage>359</fpage>
  <lpage>366</lpage>
</bibl>

<bibl id="B50">
  <title><p>High-throughput screen for aromatic hydroxylation</p></title>
  <aug>
    <au><snm>Otey</snm><fnm>C. R.</fnm></au>
    <au><snm>Joern</snm><fnm>J. M.</fnm></au>
  </aug>
  <source>Directed enzyme evolution: screening and selection methods</source>
  <publisher>Humana press</publisher>
  <editor>F. H. Arnold and G. Georgiou</editor>
  <series><title><p>Methods in Molecular Biology</p></title></series>
  <section><title><p>13</p></title></section>
  <pubdate>2003</pubdate>
  <volume>230</volume>
  <fpage>141</fpage>
  <lpage>-148</lpage>
</bibl>

<bibl id="B51">
  <title><p>Functional evolution and structural conservation in chimeric
  cytochromes P450: calibrating a structure-guided approach</p></title>
  <aug>
    <au><snm>Otey</snm><fnm>C. R.</fnm></au>
    <au><snm>Silberg</snm><fnm>J. J.</fnm></au>
    <au><snm>Voigt</snm><fnm>C. A.</fnm></au>
    <au><snm>Endelman</snm><fnm>J. B.</fnm></au>
    <au><snm>Bandara</snm><fnm>G.</fnm></au>
    <au><snm>Arnold</snm><fnm>F. H.</fnm></au>
  </aug>
  <source>Chem. Biol.</source>
  <pubdate>2004</pubdate>
  <volume>11</volume>
  <fpage>309</fpage>
  <lpage>318</lpage>
</bibl>

</refgrp>
} 


\ifthenelse{\boolean{publ}}{\end{multicols}}{}


\newpage

\section*{Figures}
\newcounter{figurenumber}

\centerline{\includegraphics[width=2.5in]{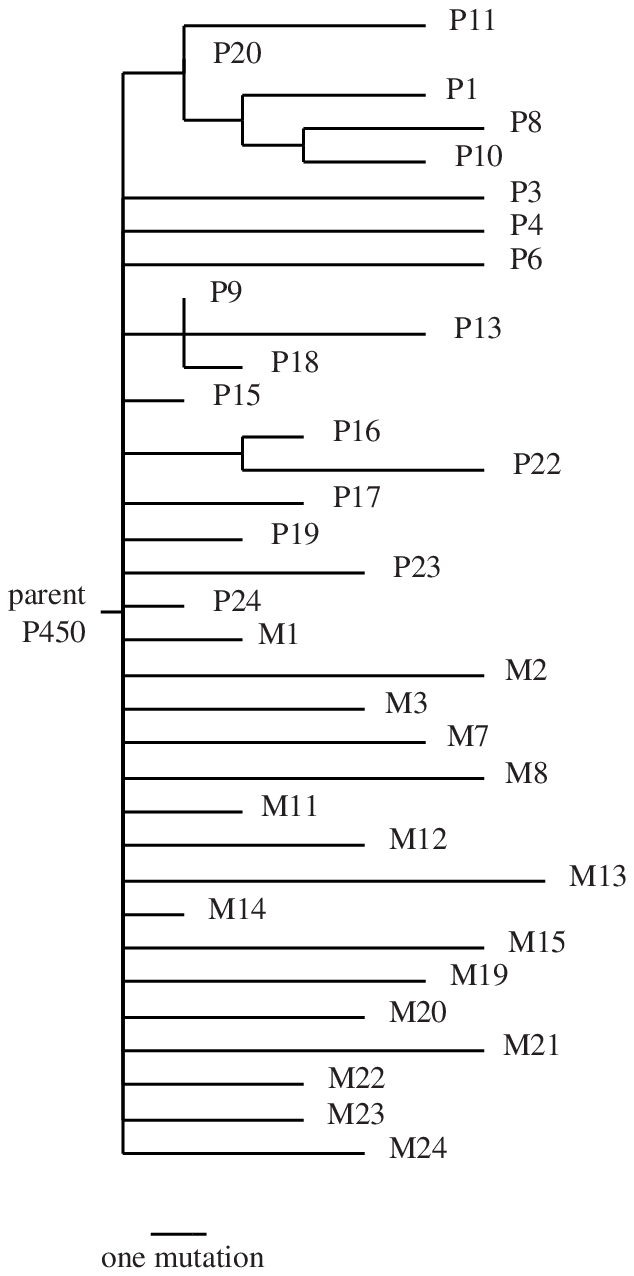}}
\subsection*{\refstepcounter{figurenumber}\label{fig:tree}Figure \arabic{figurenumber} - Phylogenetic tree of the neutrally evolved P450s}
The tree shows the relationship among the 34 neutrally evolved P450 variants examined in this study.  All of the P450s neutrally evolved from the same R1-11 parent P450.  The horizontal lengths of the branches are proportional to the number of nonsynonymous mutations, as indicated by the scale bar.  The vertical arrangement of the branches is arbitrary.
\newpage

\centerline{\includegraphics[width=5.0in]{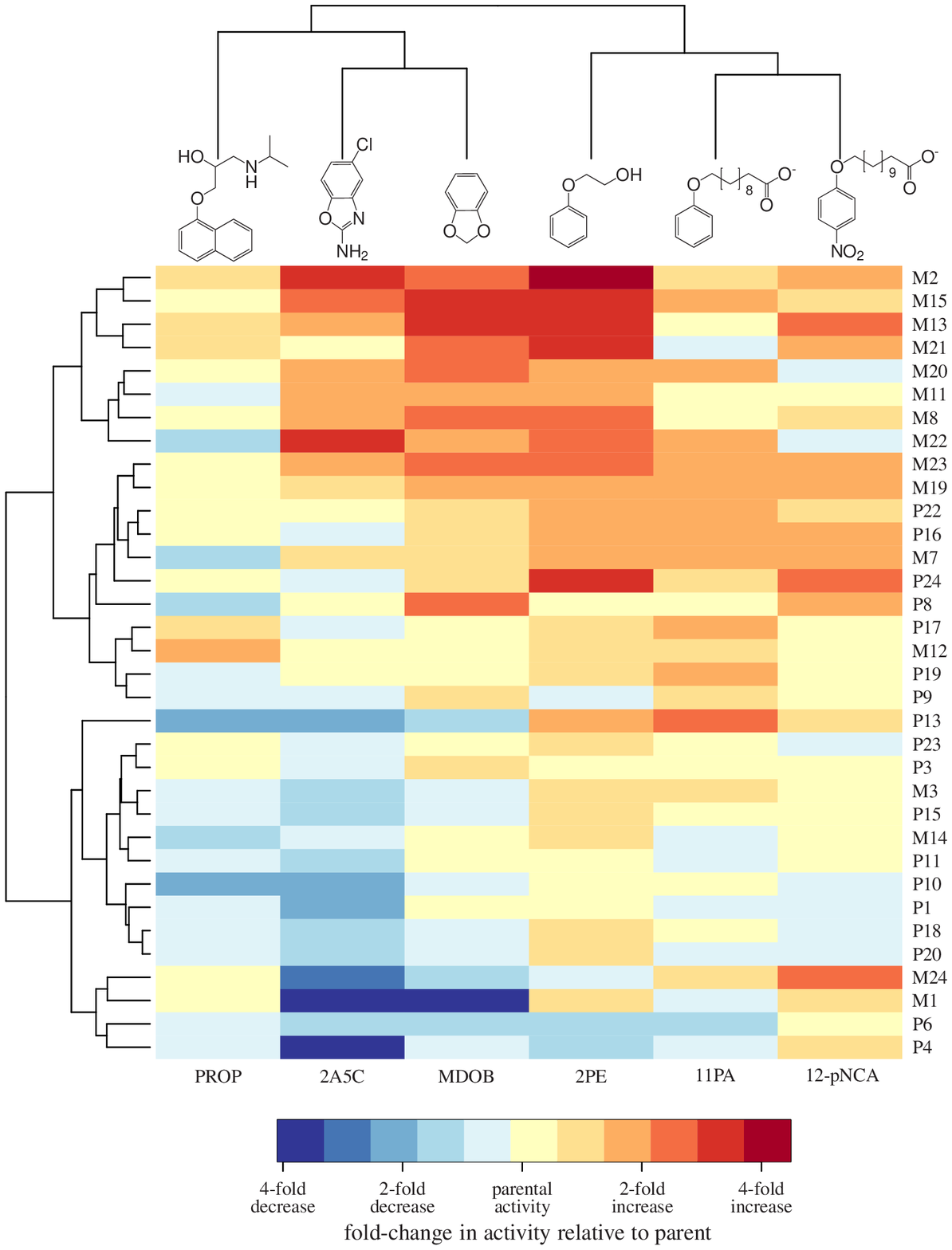}}
\subsection*{\refstepcounter{figurenumber}\label{fig:heatmap}Figure \arabic{figurenumber} - Activities of the neutrally evolved P450s on the six substrates.}
The heat map shows the fold change in activity of all 34 neutrally evolved P450 variant on all six substrates.  Each row shows the data for a different P450 variant, while each column shows the activity on a different substrate.  The fold change in activity is the ratio of the variant's activity to that of the neutral evolution parent.  Both the substrates and the P450 variants are hierarchically clustered according to the activity profiles, as shown by the dendrograms at top and left.  Substrate abbreviations: PROP - propranolol, 2A5C - 2-amino-5-chlorobenzoxazole, MDOB - 1,2-methylenedioxybenzene, 2PE - 2-phenoxyethanol, 11PA - 11-phenoxyundecanoic acid.  The standard errors for the changes in activity displayed in the heat map tend to be much smaller than the changes themselves; these errors are shown explicitly in Figure \ref{fig:foldchanges}.
\newpage

\centerline{\includegraphics[width=5in]{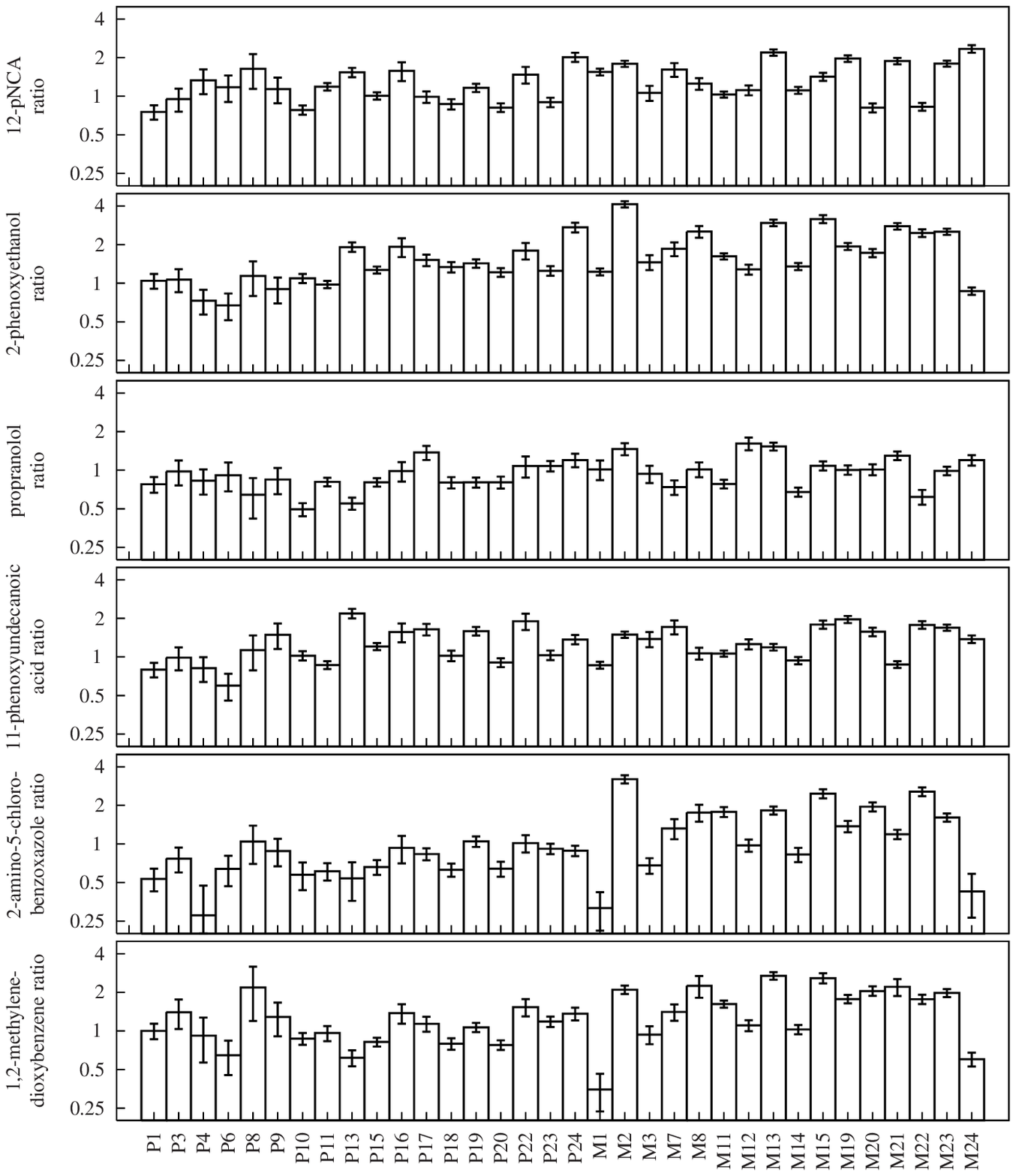}}
\subsection*{\refstepcounter{figurenumber}\label{fig:foldchanges}Figure \arabic{figurenumber} - Fold changes in P450 activities with standard errors.}
The bar graphs show the fold change in activity of all 34 neutrally evolved P450 variants on all six substrates.  This is the same data as in Figure \ref{fig:heatmap}, except these graphs also give error bars showing the standard errors in two separate measurements of the activities.  In most cases the standard errors are much smaller than the activity changes themselves.
\newpage

\includegraphics[width=6.25in]{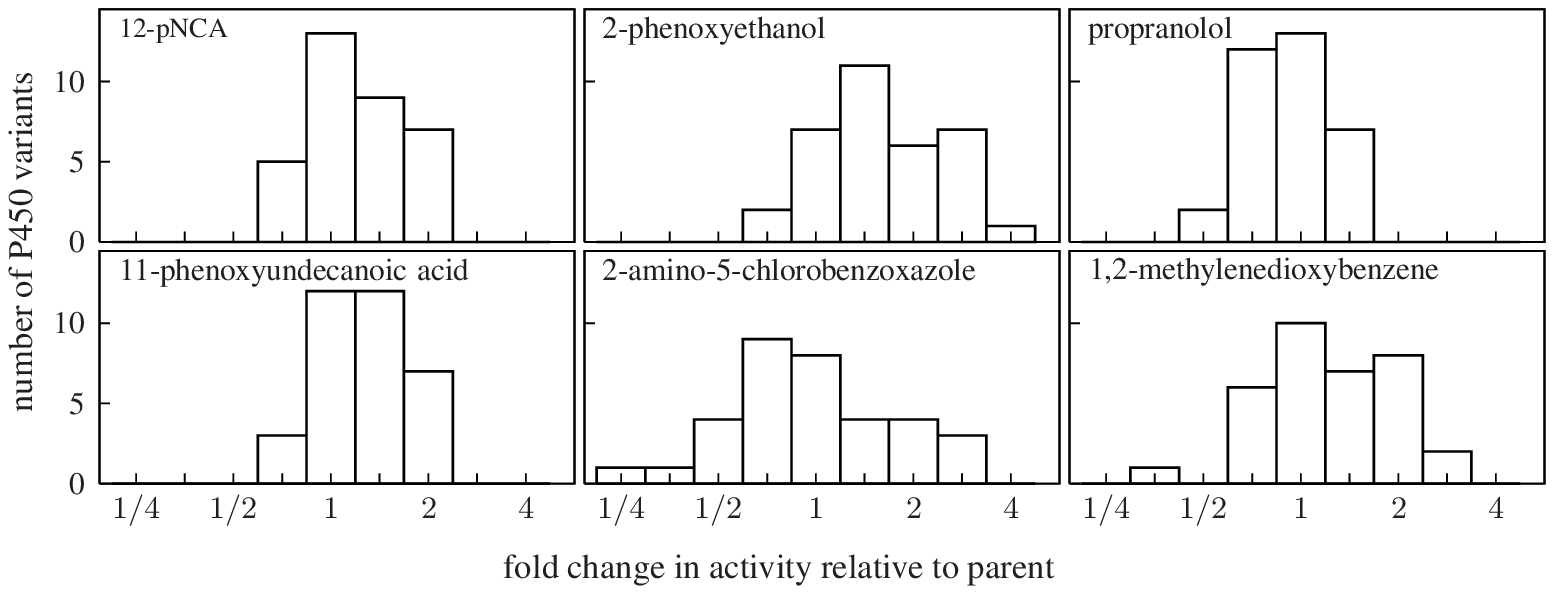}
\subsection*{\refstepcounter{figurenumber}\label{fig:changedistribution}Figure \arabic{figurenumber} - Distributions of activity changes on each of the six substrates.}
The histograms show the distributions of fold changes in activity for all 34 neutrally evolved P450 variants on each of the six substrates, with a value of one indicating that the activity is the same as the neutral evolution parent.
\newpage


\section*{Tables}
\newcounter{tablecounter}

\subsection*{\refstepcounter{tablecounter}\label{tab:pca}Table \arabic{tablecounter} - Principal component analysis of activity profiles.}
The first two principal components explain 82\% of the variance in P450 activity profiles.  The table shows the composition of these two components and the variance explained by each.  The first component contains positive contributions from all substrates and can be thought of as representing a general high catalytic ability.  The second component can be thought of as representing discrimination between fused ring substrates (PROP, 2A5C, and MDOB) and phenolic either substrates (12-pNCA, 11PA, 2PE).  Substrate abbreviations are as defined in the legend to Figure \ref{fig:heatmap}.  The analysis was performed on the logarithms of the fold changes in activity.
\par \mbox{}
\par
\mbox{
\begin{tabular}{c|cccccc}
 & 12-pNCA & 2PE  & PROP  & 11PA  & 2A5C  & MDOB \\ \hline   
PC1 (explains 62\% of variance) & 0.25 & 0.61 & 0.05 & 0.29 & 0.47 & 0.51 \\ 
PC2 (explains 20\% of variance) & -0.48 & -0.39 & 0.14 & -0.27 & 0.70 & 0.19 \\
\end{tabular}
}
\par \mbox{}
\newpage

\subsection*{\refstepcounter{tablecounter}\label{tab:mutcorrelation}Table \arabic{tablecounter} - Correlations between changes in activity and number of mutations.}
The extent of change in activity is positively correlated with the number of nonsynonymous mutations the P450 has undergone relative to the neutral evolution parent.  Each column shows the Pearson correlation between the number of nonsynonymous mutations and the absolute value of the logarithm of the fold change in activity for a different substrate, computed over all 34 P450 variants.  The final column (ALL) is the correlation among the $6 \times 34$ pooled data points for all six substrates.  The $P$-values are shown in parentheses; none of the correlations for the individual substrates are significant at a 1\% level (due to the small number of data points), but the overall correlation for all substrates is highly significant.  Substrate abbreviations are as defined in the legend to Figure \ref{fig:heatmap}.
    \par \mbox{}
    \par
    \mbox{
\begin{tabular}{r|ccccccc}
& 12-pNCA & 2PE  & PROP  & 11PA  & 2A5C  & MDOB & ALL \\ \hline   
Correlation & 0.32 (0.06) & 0.27 (0.11) & 0.06 (0.71) & 0.12 (0.48) & 0.20 (0.24) & 0.36 (0.04) & 0.22($10^{-3}$)
\end{tabular}
}
\par \mbox{}
\newpage


\section*{Additional Files}
\newcounter{addcounter}

\subsection*{\refstepcounter{addcounter}\label{add:readings}Additional file \arabic{addcounter} --- Standard curves used to determine P450 activities.}
The PDF file shows all of the standard curves used to determine the P450 concentration and enzymatic activities.  Points that were deemed to fall in the linear range, and so used to compute the standard curve slopes, are solid.  Points that were deemed outside of the linear range are empty.  Each curve shows the slopes computed for two independent replicates, and the average slope with standard error.  These average slopes were used to compute the P450 activities.

\subsection*{\refstepcounter{addcounter}\label{add:activity_data}Additional file \arabic{addcounter} --- Raw activity and sequence data.}
The text file gives the activities of the P450 variants on each of the six substrates, as measured in this study.  It also lists the number of nonsynonymous mutations (M\_NS) relative to the R1-11 neutral evolution parent sequence, as originally reported in \cite{Bloom2007c}.  Each row lists the data for a different P450 variant, and standard errors for the measured activities are shown in parentheses.  Activities are the normalized standard curve slopes as described in the Methods.

\end{bmcformat}
\end{document}